\documentclass[prb,superscriptaddress,showpacs,floatfix,tightenlines,twocolumn,10pt]{revtex4}
\usepackage{amssymb}
\usepackage{amsmath}
\usepackage{graphicx}

\begin{document}

\title{Skyrme and Wigner crystals in graphene}
\author{R. C\^{o}t\'{e}}
\affiliation{D\'{e}partement de physique and RQMP, Universit\'{e} de Sherbrooke,
Sherbrooke, Qu\'{e}bec, Canada, J1K 2R1}
\author{J.-F. Jobidon}
\affiliation{D\'{e}partement de physique and RQMP, Universit\'{e} de Sherbrooke,
Sherbrooke, Qu\'{e}bec, Canada, J1K 2R1}
\author{H. A. Fertig}
\affiliation{Department of physics, Indiana University, Bloomington, Indiana, 47405,
U.S.A.}

\begin{abstract}
At low-energy, the band structure of graphene can be approximated by two
degenerate valleys $(K,K^{\prime })$ about which the electronic spectra of
the valence and conduction bands have linear dispersion relations. An
electronic state in this band spectrum is a linear superposition of states
from the $A$ and $B$ sublattices of the honeycomb lattice of graphene. In a
quantizing magnetic field, the band spectrum is split into Landau levels
with level $N=0$ having zero weight on the $B(A)$ sublattice for the $%
K(K^{\prime })$ valley. Treating the valley index as a pseudospin and
assuming the real spins to be fully polarized, we compute the energy of
Wigner and Skyrme crystals in the Hartree-Fock approximation. We show that
Skyrme crystals have lower energy than Wigner crystals \textit{i.e.}
crystals with no pseudospin texture in some range of filling factor $\nu $
around integer fillings. The collective mode spectrum of the valley-skyrmion
crystal has three linearly-dispersing Goldstone modes in addition to the
usual phonon mode while a Wigner crystal has only one extra Goldstone mode
with a quadratic dispersion. We comment on how these modes should be
affected by disorder and how, in principle, a microwave absorption
experiment could distinguish between Wigner and Skyrme crystals.
\end{abstract}

\date{\today }
\maketitle

\section{Introduction}

For a conventional two-dimensional electron gas (2DEG)\ created in a
semiconductor heterostructure, theoretical calculations show that, in the
presence of a strong perpendicular magnetic field, a Wigner crystal (WC)\
state has lower energy then the fractional quantum Hall liquids for filling
factors $\nu \lesssim 1/6.5$\cite{lamgirvin}. Transport measurements
indicative of this electron crystallization has been reported by several
groups\cite{reviewexperimentwc}$.$ These measurements include the
observation of a strong increase in the diagonal resistivity $\rho _{xx},$
non-linear $I-V$ characteristics, and broadband noise. Another series of
experiments involving microwave absorption \cite{microwave} have also
detected a resonance in the real part of the longitudinal conductivity, $%
\sigma _{xx}\left( \omega \right) $, that has been attributed to the pinning
mode of a disordered WC. Such a resonance was observed not only at small
filling factor $\nu \lesssim 1/6.5$ in the lowest Landau level but also at
small filling factor in the higher Landau levels where the formation of a
quasiparticle solid is expected in a clean sample. In higher Landau levels,
a study of the evolution of the pinning mode with filling factor reveals
several transitions of the two-dimensional electron gas ground state from a
Wigner crystal at low $\nu $ to a series of bubble crystals with increasing
number of electrons per lattice site as $\nu $ is increased, and into a
modulated stripe state (or anisotropic Wigner crystal) near half filling\cite%
{fogler},\cite{cotebubble}.

In a conventional 2DEG, the Landau level energy spectrum is given by $%
E_{n}=\left( n+1/2\right) \hslash \omega _{c}$ where $\omega _{c}=eB/m^{\ast
}c$ is the cyclotron frequency with $m^{\ast }$ the effective mass. Each of
these levels is highly degenerate, so that a partially filled Landau level
is dominated by electron-electron interactions and is expected to enter a
crystal state. Similar physics is expected to occur in graphene. In a strong
perpendicular magnetic field, there is also a series of highly degenerate
Landau levels, with energies given by $E_{n}=sgn(n)\frac{\sqrt{2}\hslash
v_{F}}{\ell }\sqrt{\left\vert n\right\vert }$ where $v_{F}$ is the Fermi
velocity. In two recent papers, Zhang and Joglekar\cite{joglekar},\cite%
{joglekar2} have explored the possibility of Wigner crystallization in
graphene (including bubbles and stripes) in the presence of a quantizing
magnetic field. While the situations in the presence of a field are similar
for the conventional 2DEG and graphene, in the absence of a field they are
likely to be different. At low densities, electrons in the former system are
believed to form a Wigner crystal. However, graphene in zero field has a
gapless non-interacting spectrum, so that an arrangement of electrons into a
two-dimensional lattice cannot create effective potentials which localize
the individual electrons. This \textquotedblleft Klein
paradox\textquotedblright\ physics\cite{katsnelson} undermines the stability
of the Wigner crystal. In a magnetic field, however, the kinetic energy
contains a series of gaps, and the physics of the 2DEG is dominated by the
Coulomb interaction alone. A magnetic-field induced Wigner solid is thus
expected \cite{jianhui}.

In the Hartree-Fock approximation (HFA), the potential energy depends on the
effective Hartree, $H_{n}\left( \mathbf{q}\right) ,$ and Fock, $X_{n}\left( 
\mathbf{q}\right) $ interactions defined in Landau level $n.$ To an
excellent approximation\cite{goerbig}, these effective interactions for $n=0$
are identical to that of a conventional 2DEG so that the HFA phase diagram
should be the same in both systems. To develop the analogy further, we use a
pseudospin language in which the two non-equivalent valleys $K$ and $%
K^{\prime }$ of graphene (the valence and conduction bands in graphene touch
at two inequivalent points $\mathbf{K}$ and $\mathbf{K}^{\prime }=-\mathbf{K}
$ and four other points or valleys related by symmetry) are mapped to the
valley pseudospin $\left\vert \pm \right\rangle $ states. If we assume that
the real spins are completely polarized in the graphene system, then, for $%
n=0$, the HF hamiltonian for the 2DEG in graphene is identical to that of a
conventional 2DEG \textit{with zero Zeeman coupling}. For $n\neq 0$,
however, the effective interactions are not identical. On this basis, we can
expect that pseudospin skyrmion crystals are possible in graphene around
filling factor $\nu =1.$ In a conventional 2DEG, skyrmion crystals are
restricted to the lowest Landau level only but this may not be the case in
graphene. Indeed, recently Yang, Das Sarma and MacDonald have shown that
skyrmions are the lowest-energy charged excitations in graphene for Landau
levels up to $n=3$\cite{macdoyan}.

In this paper, we explore the possibility of Skyrmion crystals near integer
filling in each Landau level, and compare their stability and collective
mode properties to those of Wigner crystal states\cite{joglekar,joglekar2}.
Because there is no effective Zeeman coupling in graphene, the ground state
is generically a crystal of merons rather than of skyrmions, as it is also
the case in the conventional 2DEG\cite{breymerons}. We then compute the
collective mode dispersions in the Wigner and meron crystal phases. We show
that the approximate SU(2) symmetry of the hamiltonian leads for a Wigner
crystal to a quadratically dispersing valley-pseudospin gapless mode in
addition to the phonon mode present in the conventional 2DEG. In the meron
crystal, we find instead three new gapless modes with linear dispersions in
addition to the phonon mode. In graphene, these modes represent charge
fluctuations between the sublattices instead of spin fluctuations as in a
conventional 2DEG and so we speculate that they may be visible in microwave
absorption experiments. Each crystal structure, Wigner or Skyrme, may thus
have a unique signature in microwave absorption, in contrast to Wigner and
spin Skyrmion crystals in conventional 2DEG's where the absorption spectrum
does not distinguish between these two structures.

This paper is organized in the following way. We review in Secs. II and III
some basic properties of graphene and summarize our Hartree-Fock and
time-dependent Hartree-Fock formalism for computing the phase diagram and
collective excitations. Our numerical results are presented in Sec. IV. We
discuss in Sec. V how the collective modes that we find should be affected
by the presence of disorder and speculate on their visibility in microwave
absorption experiments. We conclude in Sec. VI.

\section{Hartree-Fock hamiltonian}

In this section, we briefly explain the Hartree-Fock formalism that we use
to compute the energy and collective excitations of the crystal states. We
start by reviewing the model hamiltonian for electrons in undoped graphene
around the Fermi energy \cite{CastroReview}.

A lattice point in graphene is given by $\mathbf{R}=n_{1}\mathbf{a}_{1}+n_{2}%
\mathbf{a}_{2}$ where $n_{1},n_{2}$ are positive or negative integers and
the primitive vectors are chosen as $\mathbf{a}_{1}=a/2\widehat{\mathbf{x}}-%
\sqrt{3}/2\widehat{\mathbf{y}}$ and $\mathbf{a}_{2}=a\widehat{\mathbf{x}}$
with the two carbon atoms in the unit cell at positions $\mathbf{r}_{1}=0$
and $\mathbf{r}_{2}=-c_{0}\widehat{\mathbf{y}}$ where $c_{0}=a/\sqrt{3}=1.42$
\AA\ is the separation between two adjacent carbon atoms. We define the
carbon atoms with basis vector $\mathbf{r}_{1}$ ($\mathbf{r}_{2}$) as part
of the $A(B)$ sublattice. The tight-binding hamiltonian for electrons in the 
$p_{z}$ orbitals of the carbon atoms is given by%
\begin{equation}
H=-t_{0}\sum_{\left\langle i,j\right\rangle }\left( a_{i}^{\dag
}b_{j}+h.c.\right) ,
\end{equation}%
where $a_{i}\left( b_{i}\right) $ is the annihilation operator for an
electron on the $A\left( B\right) $ sublattice of graphene at site $i$ and
the summation is over nearest neighbors only with hopping energy (between
different sublattices) $t_{0}\approx 2.8$ eV. In this approximation, the
dispersion relations for the valence $\left( -\right) $ and conduction $%
\left( +\right) $ bands are given by%
\begin{eqnarray}
E_{\pm }\left( \mathbf{k}\right) &=&\pm t_{0}\left[ 1+4\cos ^{2}\left( \frac{%
k_{x}a}{2}\right) \right. \\
&&\left. +4\cos \left( \frac{k_{x}a}{2}\right) \cos \left( \frac{\sqrt{3}%
k_{y}a}{2}\right) \right] ^{1/2}.  \notag
\end{eqnarray}%
For undoped graphene, the Fermi level is at energy $E=0$. With our choice of
orientation for the Bravais lattice, the positions of the two non equivalent
Dirac points are at $\mathbf{K}=-4\pi /3a\widehat{\mathbf{x}}$ and $\mathbf{K%
}^{\prime }=4\pi /3a\widehat{\mathbf{x}}$. Around each of these points in $%
\mathbf{k}-$space, the dispersion of the conduction and valence bands can be
approximated by

\begin{equation}
E\left( \mathbf{K}+\mathbf{p}\right) =E\left( \mathbf{K}^{\prime }+\mathbf{p}%
\right) \approx \pm \hslash v_{0}\left\vert \mathbf{p}\right\vert ,
\end{equation}%
where $v_{0}=3c_{0}t_{0}/2\hslash $ is the Fermi velocity. In the $\left(
A,B\right) $ basis, the hamiltonians around the Dirac points for electrons
in the conduction $\left( +\right) $ or valence $\left( -\right) $ band are
given by

\begin{equation}
H_{\pm }\left( \mathbf{p}\right) =\pm \hslash v_{0}\left( 
\begin{array}{cc}
0 & pe^{\pm i\theta _{\mathbf{p}}} \\ 
pe^{\mp i\theta _{\mathbf{p}}} & 0%
\end{array}%
\right) ,
\end{equation}%
where $\theta _{\mathbf{p}}$ is the angle between wavevector $\mathbf{p}$
and the $x$-axis.

In the presence of a transverse magnetic field $\mathbf{B}=B_{0}\widehat{%
\mathbf{z}}$, the hamiltonian is obtained by making the Peierls substitution 
$E\left( \mathbf{p}\right) \rightarrow E\left( \mathbf{p}+e\mathbf{A}%
/\hslash c\right) ,$ where $\mathbf{A}$ is the vector potential of the
magnetic field defined such that $\nabla \times \mathbf{A=B}$. In terms of
the covariant momentum $\mathbf{P}=-i\hslash \nabla +e\mathbf{A}/c=\hslash 
\mathbf{p}+e\mathbf{A}/c$, we have%
\begin{equation}
H_{\pm }\left( \mathbf{p}\right) =\pm v_{0}\left( 
\begin{array}{cc}
0 & P_{x}\pm iP_{y} \\ 
P_{x}\mp iP_{y} & 0%
\end{array}%
\right) ,
\end{equation}%
with the commutation relation 
\begin{equation}
\left[ P_{x},P_{y}\right] =-i\frac{\hslash ^{2}}{\ell ^{2}},
\end{equation}%
where $\ell =\sqrt{\hslash c/eB}$ is the magnetic length. The original
conical dispersions at Dirac points $\mathbf{K}$ and $\mathbf{K}^{\prime }$
are now split into a set of degenerate Landau levels which have quantized
energies given by%
\begin{equation}
E_{n}=sgn(n)\frac{\sqrt{2}\hslash v_{0}}{\ell }\sqrt{\left\vert n\right\vert 
},  \label{2_8}
\end{equation}%
where $n=0,\pm 1,\pm 2,...$ The wavefunctions for an electron in these
Landau levels (again in the $A,B$ basis) are given by 
\begin{eqnarray}
\left\langle \mathbf{r}|\mathbf{K}^{\prime };0,k\right\rangle &=&e^{-i%
\mathbf{K}^{\prime }\cdot \mathbf{r}}\left( 
\begin{array}{c}
h_{0,k}\left( \mathbf{r}\right) \\ 
0%
\end{array}%
\right) ,  \label{2_4} \\
\;\left\langle \mathbf{r}|\mathbf{K};0,k\right\rangle &=&e^{-i\mathbf{K}%
\cdot \mathbf{r}}\left( 
\begin{array}{c}
0 \\ 
h_{0,k}\left( \mathbf{r}\right)%
\end{array}%
\right) ,  \label{2_5}
\end{eqnarray}%
for Landau level $n=0,$ and by 
\begin{eqnarray}
\left\langle \mathbf{r}|\mathbf{K}^{\prime };n,k\right\rangle &=&\frac{1}{%
\sqrt{2}}e^{-i\mathbf{K}^{\prime }\cdot \mathbf{r}}\left( 
\begin{array}{c}
h_{\left\vert n\right\vert ,k}\left( \mathbf{r}\right) \\ 
sgn\left( n\right) h_{\left\vert n\right\vert -1,k}\left( \mathbf{r}\right)%
\end{array}%
\right) ,\;  \label{2_6} \\
\left\langle \mathbf{r}|\mathbf{K};n,k\right\rangle &=&\frac{1}{\sqrt{2}}%
e^{-i\mathbf{K}\cdot \mathbf{r}}\left( 
\begin{array}{c}
-sgn\left( n\right) h_{\left\vert n\right\vert -1,k}\left( \mathbf{r}\right)
\\ 
h_{\left\vert n\right\vert ,k}\left( \mathbf{r}\right)%
\end{array}%
\right) ,  \label{2_7}
\end{eqnarray}%
for the other levels. In the Landau gauge $\mathbf{A}=Bx\widehat{\mathbf{y}}$%
, 
\begin{eqnarray}
h_{n,k}\left( \mathbf{r}\right) &=&\left( \frac{1}{\pi \ell ^{2}L_{y}^{2}}%
\right) ^{1/4}\frac{1}{\sqrt{2^{n}n!}}e^{-iky} \\
&&\times H_{n}\left( \frac{x-k\ell }{\ell }\right) e^{-\frac{\left( x-k\ell
^{2}\right) ^{2}}{2\ell ^{2}}}.  \notag
\end{eqnarray}%
where $k=2\pi m/L_{y}$ (with $m=0,\pm 1,\pm 2,...$) and $H_{n}\left(
x\right) $ is an Hermite polynomial.

The second quantized expression for the Coulomb interaction can be written,
with the help of Eqs. (\ref{2_4}-\ref{2_7}), as

\begin{eqnarray}
U &=&\frac{1}{2S}\sum_{\mathbf{q}}V\left( \mathbf{q}\right) \int d\mathbf{r}%
\left\langle \mathbf{r}|\sigma _{1};n_{1},k_{1}\right\rangle ^{\dag }e^{i%
\mathbf{q}\cdot \mathbf{r}}\left\langle \mathbf{r}|\sigma
_{4};n_{4},k_{4}\right\rangle  \notag \\
&&\times \int d\mathbf{r}^{\prime }\left\langle \mathbf{r}^{\prime }|\sigma
_{2};n_{2},k_{2}\right\rangle ^{\dag }e^{i\mathbf{q}\cdot \mathbf{r}^{\prime
}}\left\langle \mathbf{r}^{\prime }|\sigma _{3};n_{3},k_{3}\right\rangle
\label{coulomb2} \\
&&\times c_{\sigma _{1},n_{1},k_{1}}^{\dag }c_{\sigma
_{2},n_{2},k_{2}}^{\dag }c_{\sigma _{3},n_{3},k_{3}}c_{\sigma
_{4},n_{4},k_{4}},  \notag
\end{eqnarray}%
where a summation over repeated indices is implied and where we have used $%
\sigma _{i}=\pm 1$ for the valleys at $\pm \mathbf{K}$. The Fourier
transform of the two-dimensional Coulomb interaction is $V\left( \mathbf{q}%
\right) =2\pi e^{2}/q.$ At this point, we introduce the functions $\Xi
_{n,n^{\prime }}^{\sigma ,\sigma ^{\prime }}\left( \mathbf{q}\right) $ which
we define as

\begin{eqnarray}
&&\int d\mathbf{r}\left\langle \mathbf{r}|\sigma ;n,k\right\rangle ^{\dag
}e^{i\mathbf{q}\cdot \mathbf{r}}\left\langle \mathbf{r}|\sigma ^{\prime
};n^{\prime },k^{\prime }\right\rangle \\
&\equiv &e^{\frac{i}{2}q_{x}\left( k+k^{\prime }\right) \ell ^{2}}e^{\frac{i%
}{2}(\sigma -\sigma ^{\prime })K\left( k+k^{\prime }\right) \ell ^{2}}\Xi
_{n,n^{\prime }}^{\sigma ,\sigma ^{\prime }}\left( \mathbf{q}\right) \delta
_{k,k^{\prime }-q_{y}}.  \notag
\end{eqnarray}%
These functions $\Xi _{n,n^{\prime }}^{\sigma ,\sigma ^{\prime }}\left( 
\mathbf{q}\right) $ are given by

\begin{widetext}%
\begin{eqnarray}
\Xi _{n,n^{\prime }}^{\sigma ,\sigma }\left( \mathbf{q}\right)  &=&\frac{1}{2%
}\Theta \left( \left\vert n\right\vert \right) \Theta \left( \left\vert
n^{\prime }\right\vert \right) \left[ F_{\left\vert n\right\vert ,\left\vert
n^{\prime }\right\vert }\left( \mathbf{q}\right) +sgn\left( n\right)
sgn\left( n^{\prime }\right) F_{\left\vert n\right\vert -1,\left\vert
n^{\prime }\right\vert -1}\left( \mathbf{q}\right) \right]  \\
&&+\frac{1}{\sqrt{2}}\left[ \delta _{n,0}\Theta \left( \left\vert n^{\prime
}\right\vert \right) +\delta _{n^{\prime },0}\Theta \left( \left\vert
n\right\vert \right) \right] F_{\left\vert n\right\vert ,\left\vert
n^{\prime }\right\vert }\left( \mathbf{q}\right) +\delta _{n,0}\delta
_{n^{\prime },0}F_{0,0}\left( \mathbf{q}\right) ,  \notag
\end{eqnarray}%
and%
\begin{eqnarray}
\Xi _{n,n^{\prime }}^{\sigma ,-\sigma }\left( \mathbf{q}\right)  &=&-\sigma
\frac{1}{2}\Theta \left( \left\vert n\right\vert \right) \Theta \left(
\left\vert n^{\prime }\right\vert \right) \left[ sgn\left( n\right)
F_{\left\vert n\right\vert -1,\left\vert n^{\prime }\right\vert }\left(
\mathbf{q}+\sigma 2\mathbf{K}\right) -sgn\left( n^{\prime }\right)
F_{\left\vert n\right\vert ,\left\vert n^{\prime }\right\vert -1}\left(
\mathbf{q}+\sigma 2\mathbf{K}\right) \right]  \\
&&+\sigma \frac{1}{\sqrt{2}}\left[ \delta _{n,0}\Theta \left( \left\vert
n^{\prime }\right\vert \right) sgn\left( n^{\prime }\right) F_{0,\left\vert
n^{\prime }\right\vert -1}\left( \mathbf{q}+\sigma 2\mathbf{K}\right)
-\delta _{n^{\prime },0}\Theta \left( \left\vert n\right\vert \right)
sgn\left( n\right) F_{\left\vert n\right\vert -1,0}\left( \mathbf{q}+\sigma 2%
\mathbf{K}\right) \right] ,  \notag
\end{eqnarray}%
where $\Theta \left( n\right) $ is the step function and
\begin{equation}
F_{n,n^{\prime }(n\geq n^{\prime })}\left( \mathbf{q}\right) =\left( \frac{%
n^{\prime }!}{n!}\right) ^{1/2}\left( \frac{\left( q_{y}+iq_{x}\right) \ell
}{\sqrt{2}}\right) ^{n-n^{\prime }}\exp \left( \frac{-q^{2}\ell ^{2}}{4}%
\right) L_{n^{\prime }}^{n-n^{\prime }}\left( \frac{q^{2}\ell ^{2}}{2}%
\right) ,
\end{equation}%
\end{widetext}with $F_{n,n^{\prime }(n\leq n^{\prime })}\left( \mathbf{q}%
\right) =\left[ F_{n^{\prime },n(n\geq n^{\prime })}\left( -\mathbf{q}%
\right) \right] ^{\ast }.$

In our study of crystal states, we need matrix elements of the form $\Xi
_{n,n^{\prime }}^{\sigma ,\sigma ^{\prime }}\left( \mathbf{G}\right) $ with $%
G\sim 2\pi n/a_{0}$ where $a_{0}$ is here the lattice constant of the Wigner
or Skyrme crystals. We assume that the electronic density can be made small
enough so that $a_{0}>>a$, the lattice constant of graphene. Moreover,
although the summations over $\mathbf{G}$ extend to infinity in the formulas
below, the exponential factor $\sim e^{-G^{2}\ell ^{2}/2}$ appearing in the
functions $\Xi _{n,n^{\prime }}^{\sigma ,\sigma ^{\prime }}\left( \mathbf{G}%
\right) $ makes these summations rapidly convergent if the filling factor is
not too small. We thus have an effective cutoff value $G_{\max }$ such that $%
G_{\max }<<K.$ It follows then that we can neglect the off diagonal matrix
elements $\Xi _{n,n^{\prime }}^{\sigma ,\sigma ^{\prime }\neq \sigma }\left( 
\mathbf{G}\right) $ that scatter electrons from one valley to another since
they are very small in comparison with the other terms\cite{goerbig}.
Essentially the same approximation was made in Ref. \onlinecite{joglekar}.
We also make the usual approximation of neglecting Landau level mixing. This
approximation is justified since the energy of the Landau levels are given
by Eq. (\ref{2_8}) so that the gap between the $n=0$ and $n=1$ Landau levels
is thus $\Delta _{L.L.}=\sqrt{2}\hslash v_{F}/\ell \approx 424\sqrt{B}$ K
(with $B$ in Tesla) while the Coulomb interaction energy is of the order of $%
e^{2}/\kappa \ell =130\sqrt{B}$K (with $\kappa =5$). It was recently shown
numerically that Landau level mixing is indeed negligible\cite%
{joglekar2,jianhui}.

With a Land\'{e} factor $g\approx 2$, the Zeeman gap $\Delta _{Z}=g\mu
_{B}B=1.34B$ K is however quite small in comparison with the Coulomb energy,
and the possibility of crystal states with spin as well as valley pseudospin
textures can also be considered. Previous studies of analogous bilayer 2DEG
systems\cite{bourassa} suggest that groundstates with real spin admixed are
rather fragile with respect to Zeeman coupling, and it seems unlikely that
such a textured state would be stable for this value of $g$. Our preliminary
studies of the phase diagram of the combined spin and valley pseudospin
system confirm this conclusion\cite{wluo}. In what follows we assume that
the electronic spin is fully polarized so that we need only consider the
valley degree of freedom.

In the Hartree-Fock approximation, our Hamiltonian becomes (apart from a
constant term)%
\begin{eqnarray}
H &=&N_{\phi }\sum_{\mathbf{q}}\sum_{\sigma ,\sigma ^{\prime }}H_{n}\left( 
\mathbf{q}\right) \left\langle \rho _{n}^{\sigma ,\sigma }\left( \mathbf{q}%
\right) \right\rangle \rho _{n}^{\sigma ^{\prime },\sigma ^{\prime }}\left( -%
\mathbf{q}\right)  \label{2_2} \\
&&-N_{\phi }\sum_{\sigma ,\sigma ^{\prime }}\sum_{\mathbf{q}}X_{n}\left( 
\mathbf{q}\right) \left\langle \rho _{n}^{\sigma ,\sigma ^{\prime }}\left( 
\mathbf{q}\right) \right\rangle \rho _{n}^{\sigma ^{\prime },\sigma }\left( -%
\mathbf{q}\right) ,  \notag
\end{eqnarray}%
where $N_{\phi }$ is the Landau level degeneracy and we have defined the
Hartree and Fock interactions 
\begin{eqnarray}
H_{n}\left( \mathbf{q}\right) &=&\left( \frac{e^{2}}{\kappa \ell }\right)
\left( \frac{1}{q\ell }\right) \Xi _{n}\left( \mathbf{q}\right) \Xi
_{n}\left( -\mathbf{q}\right) , \\
X_{n}\left( \mathbf{q}\right) &=&\frac{1}{N_{\phi }}\left( \frac{e^{2}}{%
\kappa \ell }\right) \sum_{\mathbf{p}}H_{n}\left( \mathbf{p}\right) e^{i%
\mathbf{p\times q}\ell ^{2}},
\end{eqnarray}%
with the form factor%
\begin{eqnarray}
\Xi _{n}\left( \mathbf{q}\right) &=&\frac{1}{2}\Theta \left( \left\vert
n\right\vert \right) \left[ F_{\left\vert n\right\vert ,\left\vert
n\right\vert }\left( \mathbf{q}\right) +F_{\left\vert n\right\vert
-1,\left\vert n\right\vert -1}\left( \mathbf{q}\right) \right]  \notag \\
&&+\delta _{n,0}F_{0,0}\left( \mathbf{q}\right) .  \label{2_9}
\end{eqnarray}%
In Eq. (\ref{2_2}), we have defined the operators 
\begin{equation}
\rho _{n}^{\sigma ,\sigma ^{\prime }}\left( \mathbf{q}\right) \equiv \frac{1%
}{N_{\phi }}\sum_{k,k^{\prime }}e^{-\frac{i}{2}q_{x}\left( k+k^{\prime
}\right) \ell ^{2}}\delta _{k,k^{\prime }+q_{y}}c_{\sigma ,n,k}^{\dagger
}c_{\sigma ^{\prime },n,k^{\prime }}.
\end{equation}%
Note that, by definition, 
\begin{equation}
\left\langle \rho _{n}^{\sigma ,\sigma }\left( \mathbf{q}=0\right)
\right\rangle =\nu _{n,\sigma },
\end{equation}%
where $\nu _{n,\sigma }=N_{n,\sigma }/N_{\phi }$ is the electronic filling
factor of the $n^{\prime }$th Landau level in the valley at $\sigma \mathbf{K%
}.$ In a crystal, the average value $\left\langle \rho _{n}^{\sigma ,\sigma
^{\prime }}\left( \mathbf{q}\right) \right\rangle $ is non zero only for $%
\mathbf{q}=\mathbf{G}$, a reciprocal lattice vector.

The ground-state energy per electron in Landau level $n$ is simply 
\begin{eqnarray}
\frac{E}{N_{n}} &=&\frac{1}{2\nu _{n}}\sum_{\mathbf{q}}\sum_{\sigma ,\sigma
^{\prime }}H_{n}\left( \mathbf{q}\right) \left\langle \rho _{n}^{\sigma
,\sigma }\left( \mathbf{q}\right) \right\rangle \left\langle \rho
_{n}^{\sigma ^{\prime },\sigma ^{\prime }}\left( -\mathbf{q}\right)
\right\rangle  \notag \\
&&-\frac{1}{2\nu _{n}}\sum_{\sigma ,\sigma ^{\prime }}\sum_{\mathbf{q}%
}X_{n}\left( \mathbf{q}\right) \left\vert \left\langle \rho _{n}^{\sigma
,\sigma ^{\prime }}\left( \mathbf{q}\right) \right\rangle \right\vert ^{2}.
\label{energiehf2}
\end{eqnarray}%
From Eq. (\ref{2_9}), we see that the form factor $\Xi _{0}\left( \mathbf{q}%
\right) $ for Landau level $n=0$ is exactly the same as for a 2DEG in a
semiconductor quantum well or heterostructure. It follows that the phase
diagram for graphene at low filling factor will be closely related to that
of a conventional 2DEG with vanishing Zeeman gap\cite{cotecp3}.

It is very useful to map the valley degree of freedom into a pseudospin $1/2$%
. Our convention is that a $K$ state is pseudospin up $\left( +\right) $
while $K^{\prime }$ is pseudospin down $\left( -\right) .$ In this language,
the components of the pseudospin vector density $\mathbf{P}_{n}\left( 
\mathbf{q}\right) =P_{n,x}\left( \mathbf{q}\right) \widehat{\mathbf{x}}%
+P_{n,y}\left( \mathbf{q}\right) \widehat{\mathbf{y}}+P_{n,z}\left( \mathbf{q%
}\right) \widehat{\mathbf{z}}$ are given by $\mathbf{\ }$%
\begin{eqnarray}
P_{n,x}\left( \mathbf{q}\right) &=&\frac{\rho _{n}^{+,-}\left( \mathbf{q}%
\right) +\rho _{n}^{-,+}\left( \mathbf{q}\right) }{2}, \\
P_{n,y}\left( \mathbf{q}\right) &=&\frac{\rho _{n}^{+,-}\left( \mathbf{q}%
\right) -\rho _{n}^{-,+}\left( \mathbf{q}\right) }{2i}, \\
P_{n,z}\left( \mathbf{q}\right) &=&\frac{\rho _{n}^{+,+}\left( \mathbf{q}%
\right) -\rho _{n}^{-,-}\left( \mathbf{q}\right) }{2},
\end{eqnarray}%
while the total density is given by%
\begin{equation}
\rho _{n}\left( \mathbf{q}\right) =\rho _{n}^{+,+}\left( \mathbf{q}\right)
+\rho _{n}^{-,-}\left( \mathbf{q}\right) .
\end{equation}%
In this language, Eq. (\ref{energiehf2}) becomes

\begin{eqnarray}
\frac{E}{N_{n}} &=&\frac{1}{4\nu _{n}}\left( \frac{e^{2}}{\kappa \ell }%
\right) \sum_{\mathbf{q}}\Upsilon _{n}\left( \mathbf{q}\right) \left\langle
\rho _{n}\left( -\mathbf{q}\right) \right\rangle \left\langle \rho
_{n}\left( \mathbf{q}\right) \right\rangle  \label{hfpseudospin} \\
&&+\frac{1}{\nu _{n}}\left( \frac{e^{2}}{\kappa \ell }\right) \sum_{\mathbf{q%
}}J_{n}\left( \mathbf{q}\right) \left[ \left\langle \mathbf{P}_{n}\left( -%
\mathbf{q}\right) \right\rangle \cdot \left\langle \mathbf{P}_{n}\left( 
\mathbf{q}\right) \right\rangle \right] ,  \notag
\end{eqnarray}%
with%
\begin{eqnarray}
\Upsilon _{n}\left( \mathbf{q}\right) &=&2H_{n}\left( \mathbf{q}\right)
-X_{n}\left( \mathbf{q}\right) , \\
J_{n}\left( \mathbf{q}\right) &=&-X_{n}\left( \mathbf{q}\right)
=-\int_{0}^{\infty }dye^{-y^{2}/2}J_{0}\left( yq\ell \right) ,
\end{eqnarray}%
where $J_{0}\left( x\right) $ is a Bessel function. For example, the liquid
state at $\nu =1$ has an energy given by 
\begin{equation}
\frac{E}{N_{n}}=-\left[ \frac{1}{4}X_{n}\left( 0\right) +X_{n}\left(
0\right) \left\vert \left\langle \mathbf{P}_{n}\left( 0\right) \right\rangle
\right\vert ^{2}\right] ,
\end{equation}%
and $\left\vert \left\langle \mathbf{P}_{n}\left( 0\right) \right\rangle
\right\vert =1/2$. (We have taken into account the positive background to
cancel the divergence of $H_{n}\left( 0\right) $). It follows that this
liquid state is fully pseudospin polarized but the direction of polarization
is arbitrary. This is also true for the crystal states (see below) i.e. our
hamiltonian has an $SU(2)$ symmetry. We deduce that both states support a
pseudospin wave Goldstone mode with a $q^{2}$ dispersion at long wavelength.

We remark that, in view of Eqs. (\ref{2_4})-(\ref{2_5}), the pseudospin
degree of freedom is equivalent to the sublattice degree of freedom for
Landau level $n=0$. This is not true, however, for other values of $n$.

\section{Single and two-particle Green's functions}

The average values $\left\langle \rho _{n}^{\sigma ,\sigma ^{\prime }}\left( 
\mathbf{B}\right) \right\rangle $\ describing the crystal states can be
extracted from the Matsubara single-particle Green's function $G_{n}$ which
is defined by (with $X=k\ell ^{2}$) 
\begin{equation}
G_{n}^{\sigma ,\sigma ^{\prime }}\left( X,X^{\prime },\tau \right)
=-\left\langle Tc_{\sigma ,n,X}\left( \tau \right) c_{\sigma ^{\prime
},n^{\prime },X^{\prime }}^{\dagger }\left( 0\right) \right\rangle .
\end{equation}%
Its Fourier transform is 
\begin{eqnarray}
G_{n}^{\sigma ,\sigma ^{\prime }}\left( \mathbf{G,}\tau =0^{-}\right) &=&%
\frac{1}{N_{\phi }}\sum_{X,X^{\prime }}e^{-\frac{i}{2}G_{x}\left(
X+X^{\prime }\right) } \\
&&\times \delta _{X,X^{\prime }-G_{y}l_{\bot }^{2}}G_{n}^{\sigma ,\sigma
^{\prime }}\left( X,X^{\prime },\tau \right) ,  \notag
\end{eqnarray}%
so that 
\begin{equation}
\left\langle \rho _{n}^{\sigma ^{\prime },\sigma }\left( \mathbf{G}\right)
\right\rangle =G_{n}^{\sigma ,\sigma ^{\prime }}\left( \mathbf{G,}\tau
=0^{-}\right) .
\end{equation}

The equation of motion for the Green's function in the Matsubara formalism
is obtained by using the Heisenberg equation 
\begin{equation}
\hslash \frac{\partial }{\partial \tau }\left( \ldots \right) =\left[
H_{HF}-\mu N,\left( \ldots \right) \right] ,
\end{equation}%
and is given by 
\begin{eqnarray}
&&\hslash \left( i\omega _{n}-\mu \right) G_{n}^{\sigma ,\sigma ^{\prime
}}\left( \mathbf{G},\omega _{n}\right)  \label{2_10} \\
&&-\sum_{\sigma ^{\prime \prime }}\sum_{\mathbf{G}^{\prime }}F_{\mathbf{G},%
\mathbf{G}^{\prime }}^{\sigma ,\sigma ^{\prime \prime }}e^{-i\mathbf{G}%
\times \mathbf{G}^{\prime }\ell ^{2}/2}G_{n}^{\sigma ^{\prime \prime
},\sigma ^{\prime }}\left( \mathbf{G}^{\prime },\omega _{n}\right)  \notag \\
&=&\hslash \delta _{\mathbf{G},0}\delta _{\sigma ,\sigma ^{\prime }},  \notag
\end{eqnarray}%
where $\omega _{n}$ is a fermionic Matsubara frequency and the matrix
elements $F_{\mathbf{G},\mathbf{G}^{\prime }}^{\sigma ,\sigma ^{\prime }}$
are given by 
\begin{eqnarray}
F_{n}^{\sigma ,\sigma ^{\prime }}\left( \mathbf{G},\mathbf{G}^{\prime
}\right) &=&H_{n}\left( \mathbf{G}-\mathbf{G}^{\prime }\right) \sum_{\sigma
^{\prime }}\left\langle \rho _{n}^{\sigma ^{\prime },\sigma ^{\prime
}}\left( \mathbf{G}-\mathbf{G}^{\prime }\right) \right\rangle \delta
_{\sigma ,\sigma ^{\prime }}  \notag \\
&&-X_{n}\left( \mathbf{G}-\mathbf{G}^{\prime }\right) \left\langle \rho
_{n}^{\sigma ^{\prime },\sigma }\left( \mathbf{G}-\mathbf{G}^{\prime
}\right) \right\rangle .
\end{eqnarray}

To find the order parameters $\left\langle \rho _{n}^{\sigma ^{\prime
},\sigma }\left( \mathbf{G}\right) \right\rangle ,$ we solve the
Hartree-Fock equation of motion numerically by an iterative method. The
procedure is described in detail in Ref. \onlinecite{cotemethode}.

In order to compute the collective excitations, we define the $16$ response
functions%
\begin{eqnarray}
\chi _{n}^{a,b,c,d}\left( \mathbf{q},\mathbf{q}^{\prime };\tau \right)
&=&-N_{\phi }\left\langle T\rho _{n}^{a,b}\left( \mathbf{q,}\tau \right)
\rho _{n}^{c,d}\left( -\mathbf{q}^{\prime },0\right) \right\rangle  \notag \\
&&+N_{\phi }\left\langle \rho _{n}^{a,b}\left( \mathbf{q}\right)
\right\rangle \left\langle \rho _{n}^{c,d}\left( -\mathbf{q}^{\prime
}\right) \right\rangle ,
\end{eqnarray}%
where $a,b,c,d$ are valley indices. For a crystal, $\mathbf{q=k+G}$ and $%
\mathbf{q=k+G}^{\prime },$ where $\mathbf{k}$ is a vector in the first
Brillouin zone. In the generalized random-phase approximation (GRPA), the
equation of motion for $\chi _{n}^{a,b,c,d}\left( \mathbf{q},\mathbf{q}%
^{\prime };i\Omega _{n}\right) $ is given by%
\begin{equation}
\sum_{\mathbf{q}^{\prime \prime }}\left[ i\Omega _{n}I-\frac{1}{\hslash }%
F\left( \mathbf{q,q}^{\prime \prime }\right) \right] \chi \left( \mathbf{q}%
^{\prime \prime },\mathbf{q}^{\prime };i\Omega _{n}\right) =D\left( \mathbf{%
q,q}^{\prime }\right) ,  \label{2_3}
\end{equation}%
where $I$ is the $4N_{\mathbf{q}}\times N_{\mathbf{q}}$ unit matrix with $N_{%
\mathbf{q}}$ the number of vectors $\mathbf{G}$ kept in the calculation and
we have defined the matrices:%
\begin{equation}
\chi =\left( 
\begin{array}{cccc}
\chi _{+,+,+,+} & \chi _{+,+,-,+} & \chi _{+,+,+,-} & \chi _{+,+,-,-} \\ 
\chi _{+,-,+,+} & \chi _{+,-,-,+} & \chi _{+,-,+,-} & \chi _{+,-,-,-} \\ 
\chi _{-,+,+,+} & \chi _{-,+,-,+} & \chi _{-,+,+,-} & \chi _{-,+,-,-} \\ 
\chi _{-,-,+,+} & \chi _{-,-,-,+} & \chi _{-,-,+,-} & \chi _{-,-,-,-}%
\end{array}%
\right) ,  \label{2_11}
\end{equation}%
and 
\begin{eqnarray}
F\left( \mathbf{q,q}^{\prime }\right) &=&U\left( \mathbf{q,q}^{\prime
}\right) \\
&&-\sum_{\mathbf{q}^{\prime \prime }}D\left( \mathbf{q,q}^{\prime \prime
}\right) \left( \widetilde{H}\left( \mathbf{q}^{\prime \prime },\mathbf{q}%
^{\prime }\right) -\widetilde{X}\left( \mathbf{q}^{\prime \prime },\mathbf{q}%
^{\prime }\right) \right) ,
\end{eqnarray}%
with

\begin{widetext}

\begin{equation}
D=\left(
\begin{array}{cccc}
2i\alpha \left\langle \rho _{n}^{+,+}\right\rangle  & -\left\langle \rho
_{n}^{-,+}\right\rangle \gamma  & \left\langle \rho _{n}^{+,-}\right\rangle
\gamma ^{\ast } & 0 \\
-\left\langle \rho _{n}^{+,-}\right\rangle \gamma  & \left\langle \rho
_{n}^{+,+}\right\rangle \gamma ^{\ast }-\left\langle \rho
_{n}^{-,-}\right\rangle \gamma  & 0 & \left\langle \rho
_{n}^{+,-}\right\rangle \gamma ^{\ast } \\
\left\langle \rho _{n}^{-.+}\right\rangle \gamma ^{\ast } & 0 & \left\langle
\rho _{n}^{-,-}\right\rangle \gamma ^{\ast }-\left\langle \rho
_{n}^{+,+}\right\rangle \gamma  & -\left\langle \rho _{n}^{-,+}\right\rangle
\gamma  \\
0 & \left\langle \rho _{n}^{-,+}\right\rangle \gamma ^{\ast } &
-\left\langle \rho _{n}^{+,-}\right\rangle \gamma  & 2i\alpha \left\langle
\rho _{n}^{-,-}\right\rangle
\end{array}%
\right) ,
\end{equation}%
and%
\begin{equation}
U=\left(
\begin{array}{cccc}
\begin{array}{c}
-2i\alpha H\left[ \left\langle \rho _{n}^{+,+}\right\rangle +\left\langle
\rho _{n}^{-,-}\right\rangle \right]  \\
+2i\alpha X\left\langle \rho _{n}^{+,+}\right\rangle
\end{array}
& -X\left\langle \rho _{n}^{+,-}\right\rangle \gamma  & X\left\langle \rho
_{n}^{-,+}\right\rangle \gamma ^{\ast } & 0 \\
-X\left\langle \rho _{n}^{-,+}\right\rangle \gamma  &
\begin{array}{c}
-2i\alpha H\left[ \left\langle \rho _{n}^{+,+}\right\rangle +\left\langle
\rho _{n}^{-,-}\right\rangle \right]  \\
+X\left\langle \rho _{n}^{+,+}\right\rangle \gamma ^{\ast }-X\left\langle
\rho _{n}^{-,-}\right\rangle \gamma
\end{array}
& 0 & X\left\langle \rho _{n}^{-,+}\right\rangle \gamma ^{\ast } \\
X\left\langle \rho _{n}^{+,-}\right\rangle \gamma  & 0 &
\begin{array}{c}
-2i\alpha H\left[ \left\langle \rho _{n}^{+,+}\right\rangle +\left\langle
\rho _{n}^{-,-}\right\rangle \right]  \\
+X\left\langle \rho _{n}^{-,-}\right\rangle \gamma ^{\ast }-X\left\langle
\rho _{n}^{+,+}\right\rangle \gamma
\end{array}
& -X\left\langle \rho _{n}^{+,-}\right\rangle \gamma  \\
0 & X\left\langle \rho _{n}^{+,-}\right\rangle \gamma ^{\ast } &
-X\left\langle \rho _{n}^{-,+}\right\rangle \gamma  &
\begin{array}{c}
-2i\alpha H\left[ \left\langle \rho _{n}^{+,+}\right\rangle +\left\langle
\rho _{n}^{-,-}\right\rangle \right]  \\
+2i\alpha X\left\langle \rho _{n}^{-,-}\right\rangle
\end{array}%
\end{array}%
\right).
\end{equation}%
\end{widetext}

In these equations, we adopt the conventions that $\left\langle \rho
^{a,b}\right\rangle =\left\langle \rho ^{a,b}\left( \mathbf{q-q}^{\prime
}\right) \right\rangle ,$ $H\left\langle \rho ^{a,b}\right\rangle $ stands
for $H_{n}\left( \mathbf{q-q}^{\prime }\right) \left\langle \rho
^{a,b}\left( \mathbf{q-q}^{\prime }\right) \right\rangle ,\gamma =e^{-i%
\mathbf{q}\times \mathbf{q}^{\prime }\ell ^{2}/2},$ and $\alpha =\sin \left( 
\frac{\mathbf{q}\times \mathbf{q}^{\prime }\ell ^{2}}{2}\right) .$ The
Hartree and Fock interaction matrices are given by%
\begin{equation}
\widetilde{H}=\left( 
\begin{array}{cccc}
H_{n}\left( \mathbf{q}\right) & 0 & 0 & H_{n}\left( \mathbf{q}\right) \\ 
0 & 0 & 0 & 0 \\ 
0 & 0 & 0 & 0 \\ 
H_{n}\left( \mathbf{q}\right) & 0 & 0 & H_{n}\left( \mathbf{q}\right)%
\end{array}%
\right) \delta _{\mathbf{q},\mathbf{q}^{\prime }},
\end{equation}%
and 
\begin{equation}
\widetilde{X}=\left( 
\begin{array}{cccc}
X_{n}\left( \mathbf{q}\right) & 0 & 0 & 0 \\ 
0 & X_{n}\left( \mathbf{q}\right) & 0 & 0 \\ 
0 & 0 & X_{n}\left( \mathbf{q}\right) & 0 \\ 
0 & 0 & 0 & X_{n}\left( \mathbf{q}\right)%
\end{array}%
\right) \delta _{\mathbf{q},\mathbf{q}^{\prime }}.
\end{equation}%
Once the Hartree-Fock densities $\left\langle \rho ^{a,b}\right\rangle $ are
calculated for the crystal state considered, the response functions can be
computed using Eq. (\ref{2_3}). The collective excitations appear as poles
of these response functions. To derive the dispersion relations, we follow
the poles in the response functions as the wave vector $\mathbf{k}$ is
varied within the first Brillouin zone. We consider only the low-energy
modes in the present work. The response functions also have higher-energy
modes corresponding to more localized excitations.

The various response functions in Eq. (\ref{2_11}) can be combined in an
obvious way to give the pseudospin response functions $\chi _{\rho _{n},\rho
_{n}},\chi _{S_{x},S_{x}},\chi _{S_{y},S_{y}},$ and $\chi _{S_{z},S_{z}}.$

\section{Phase diagram and collective modes}

In the absence of Zeeman coupling, each Landau level has a fourfold
degeneracy (the valley degeneracy combined with the usual spin doublet). For
undoped graphene, the $n=0$ Landau level multiplet is half-filled. We use
the notation $\nu _{n}\in \left[ 0,4\right] $ for the filling factor of each
Landau level multiplet. The total filling factor is thus given by $\nu
=4n-2+\nu _{n}.$ In this work, we assume a finite Zeeman coupling but
neglect any mixing of Landau level with different spins so that the phase
diagram for $\nu _{n}\in \left[ 0,2\right] $ is identical with that for $\nu
_{n}\in \left[ 2,4\right] .$ Without lost of generality, we consider $\nu
_{n}\in \left[ 0,2\right] $ from now on.

Our procedure for solving Eq. (\ref{2_10}) does not allow us to find the
absolute ground state of the 2DEG for a given filling factor. Instead, we
have to be content with comparing the energy of different phases and finding
the lowest one amongst them. In this study, we focus on the crystal states
and more specifically on the Skyrme crystals. The filling factor $\nu _{n}$
is that of the partially filled Landau level and all filled levels below $n$
are assumed inert. This procedure is valid when Landau level mixing is
small, provided one considers only intra-Landau level excitations\cite%
{jianhui,iyengar}. We consider the following states in each Landau level $n:$

\begin{enumerate}
\item Electron bubble crystal (eBCn). A triangular lattice with $N_{c}$
electrons per unit cell and filling factor $\nu _{n}<0.5.$ More precisely,
bubbles are maximal density droplets\cite{fogler,cotebubble}.

\item Hole bubble crystal (hBCn). A triangular lattice with $N_{c}$ holes
per unit cell and filling factor $0.5<\nu _{n}<1.$ The lattice constant $%
a_{0}$ of such a crystal is determined by the relation $2\pi n_{h}\ell
^{2}=\left\vert \nu _{n}-1\right\vert $ where $n_{h}=N_{c}/\varepsilon
a_{0}^{2}$ is the hole density with $N_{c}$ the number of holes in the
bubbles and $\varepsilon =\sqrt{3}/2$ for a triangular lattice. Note that we
find both the hBCn state and the meron crystal considered below to be lower
in energy that the eBCn groundstate assumed in Ref. \onlinecite{joglekar} in
the same range of filling factors.

\item Meron crystal (MC). A square lattice with four merons of charge $-e/2$
(if $\nu _{n}<1$) or $e/2$ (if $\nu _{n}>1$) per unit cell, equally spaced
and assembled in a checkerboard configuration. The $z$ component of the
pseudospin and the vorticities alternate from one site to the next and two
of the merons in the unit cell have a global pseudospin phase in the $x-y$
plane which is opposite to the two others. In semiconductor 2DEG's, this
configuration is found for the (spin) skyrmion crystal when the Zeeman
energy is zero\cite{breymerons}. This crystal is represented in Fig. \ref%
{fig1}.

\item Meron pair crystal (MPC). A triangular lattice with four merons per
unit cell. At each lattice site, two merons with the same value of $P_{z}$
and vorticities but opposite values of the global phase are coupled together
so that the pseudospins rotate by $4\pi $ on a path encircling the two meron
pairs. This configuration is represented in Fig. \ref{fig2}. The merons are
not equally spaced. The possibility for skyrmions of opposite phases to form
pairs was considered in Ref. \onlinecite{Nazarov}. The MC and MPC phases are
in competition with each other and their energies are very close. We remark
that, in the absence of an equivalent Zeeman coupling, $\Delta _{Z}$, a
single skyrmion should have a size comparable to that of the sample size. In
a lattice, this causes a strong interaction between skyrmions that leads to
a lattice of merons even when the skyrmion filling factor $\left\vert \nu
_{n}-1\right\vert \rightarrow 0$. It is important to notice that the limits $%
\Delta _{Z}\rightarrow 0$ and $\left\vert \nu _{n}-1\right\vert \rightarrow
0 $ do not commute\cite{breymerons}. If we were to choose $\left\vert \nu
_{n}-1\right\vert \rightarrow 0$ first and then $\Delta _{Z}\rightarrow 0$,
we would find instead a Skyrme crystal\cite{breymerons}.
\end{enumerate}

For the numerical calculations, we consider a filling factor $\nu _{n}\in $ $%
\left[ 0.1,0.9\right] $. For $\nu _{n}<0.1$ or $\left\vert \nu
_{n}-1\right\vert <0.1$, the number of reciprocal lattice vectors needed in
the calculation becomes very large and we do not get good convergence. This
is due to the fact that the size in real space of the quasiparticles
(electrons for $\nu _{n}<0.1$ or holes or skyrmions for $\left\vert \nu
_{n}-1\right\vert <0.1$) decreases so that more wavevectors are needed to
describe them. Also, the hamiltonian has electron-hole symmetry around $\nu
_{n}=1$ so that the sequence of phase transitions found for $\nu _{n}>1$ is
the mirror image of that for $\nu _{n}<1$ with particles replaced by
anti-particles. For example, the couterpart of the phase eBC1 at $\nu
_{n}=0.2$ is a hBC1 at $\nu _{n}=1.8$ with a filling of holes given by $\nu
_{n,h}=2.0-1.8=0.2.$ Similarly, the counterpart of a crystal of merons (with
charge $-e/2$) at $\nu _{n}=0.8$ (with a filling of merons given by $\nu
_{n,m}=1.0-0.8=0.2$) is a crystal of anti-merons (charge $+e/2$) at $\nu
_{n}=1.2$ with a filling of anti-merons given by $\nu _{n,am}=1.2-1.0=0.2$.

\begin{figure}[tbph]
\includegraphics[scale=1]{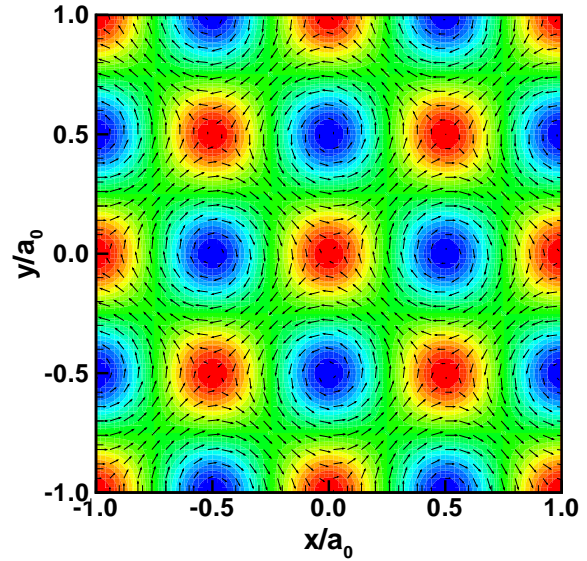}
\caption{Pseudospin texture in a meron crystal at filling factor $\protect%
\nu _{0}=0.8$ in Landau level $n=0$. The crystal has four merons per unit
cell. In each unit cell, two merons with the same vorticity have opposite
phases as explained in the text. Contours (ranging from $-0.5$ to $0.5$)
indicate the $z$ component of the pseudospin with dark regions corresponding
to positive values.}
\label{fig1}
\end{figure}

\begin{figure}[tbph]
\includegraphics[scale=1]{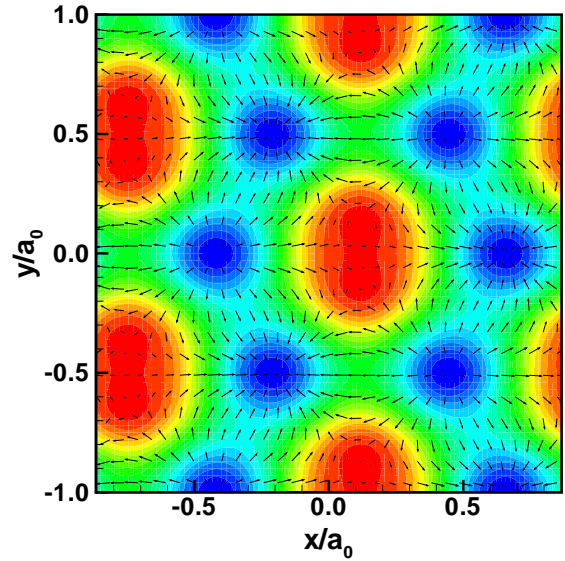}
\caption{Pseudospin texture in a meron pair crystal at filling factor $%
\protect\nu _{0}=0.8$ in Landau level $n=0.$ The lattice is triangular and
there are four merons per unit cell. Merons are bound in pairs with same
value of $P_{z}$ and vorticities but opposite phases at each lattice site.
Contours (ranging from $-0.5$ to $0.5$) indicate the $z$ component of the
pseudospin with dark regions corresponding to positive values.}
\label{fig2}
\end{figure}

\begin{figure}[tbph]
\includegraphics[scale=1]{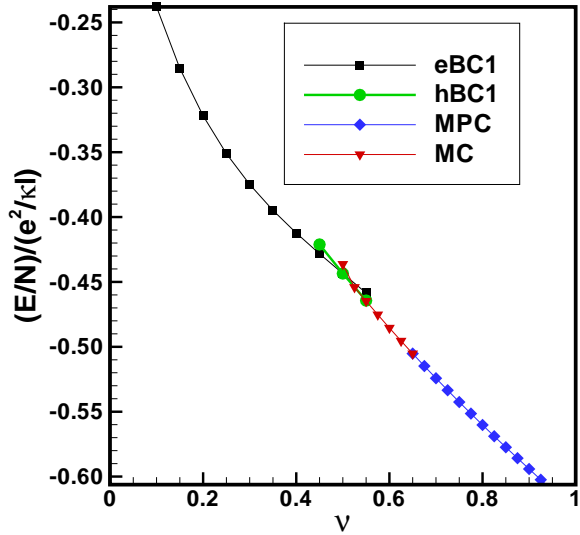}
\caption{Hartree-Fock energy per electron as a function of filling factor
for various crystal phases in Landau level $n=0.$}
\label{fig3}
\end{figure}

We show in Fig. \ref{fig3} the energies of different phases of the 2DEG in
graphene for Landau level $n=0.$ We find the following sequence: eBC1 for $%
\nu _{0}\in \left[ 0.1,0.5\right] ,$ hBC1 for $\nu _{0}\in \left[ 0.5,0.55%
\right] ,$ MC for $\nu _{0}\in \left[ 0.55,0.65\right] ,$ and MCP for $\nu
_{0}\in \left[ 0.65,0.9\right] $. As noted above, this sequence of
transitions is the same as that calculated for a 2DEG\ in GaAs-AlGaAs
quantum wells in the absence of Zeeman coupling because the effective
interactions $H_{0}\left( \mathbf{q}\right) $ and $X_{0}\left( \mathbf{q}%
\right) $ are the same in both cases. To determine this sequence, we not
only find the state with the lowest energy but also compute the collective
mode spectrum in order to check that the crystal is stable. For the MC and
MPC where the difference in energy is close to our numerical accuracy, the
stability criteria allows us to find the correct ground state.

\begin{figure}[tbph]
\includegraphics[scale=1]{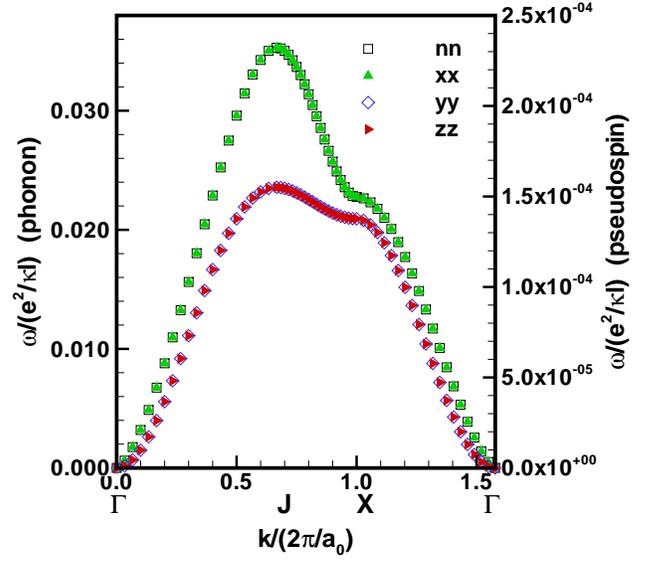}
\caption{Dispersion relation of the two Goldstone modes of the eBC1 state at 
$\protect\nu _{0}=0.2$ in Landau level $n=0.$ The dispersion is plotted
along the irreducible Brillouin zone of the triangular lattice. The left
(right) $y$ axis gives the phonon (pseudospin) frequency. The phonon mode
has its biggest weight in $\protect\chi _{\protect\rho _{n},\protect\rho %
_{n}}$ and $\protect\chi _{S_{x},S_{x}}$ while the pseudospin wave mode has
its biggest weight in $\protect\chi _{S_{y},S_{y}}$ and $\protect\chi %
_{S_{z},S_{z}}$}
\label{fig4}
\end{figure}

\begin{figure}[tbph]
\includegraphics[scale=1]{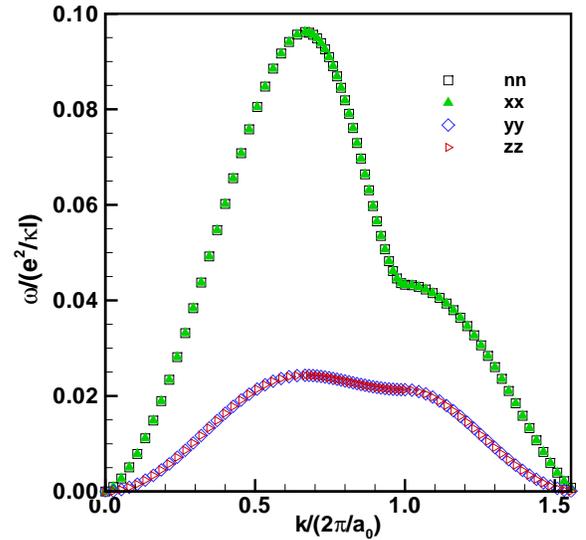}
\caption{Dispersion relation of the two Goldstone modes of the hBC1 state at 
$\protect\nu _{1}=0.55$ in Landau level $n=1.$ The dispersion is plotted
along the irreducible Brillouin zone of the triangular lattice. The phonon
mode has its biggest weight in $\protect\chi _{\protect\rho _{n},\protect%
\rho _{n}}$ and $\protect\chi _{S_{x},S_{x}}$ while the pseudospin wave mode
has its biggest weight in $\protect\chi _{S_{y},S_{y}}$ and $\protect\chi %
_{S_{z},S_{z}}$}
\label{fig5}
\end{figure}

The eBC1 phase is fully pseudospin polarized but its energy is independent
of the orientation of the pseudospins. The crystal thus has a full SU(2)
symmetry for the pseudospin. The dispersion relations of the two Goldstone
modes of this crystal is given in Fig. \ref{fig4}. The dispersion is plotted
along the path $\Gamma -J-X-\Gamma ,$corresponding to the wave vectors $%
\left( k_{x},k_{y}\right) =\left( 0,0\right) ,\left( 2\pi _{0}/a\right)
\left( 1/\sqrt{3},1/3\right) ,\left( 2\pi /a_{0}\right) \left( 1/\sqrt{3}%
,0\right) ,\left( 0,0\right) $. The wave vector $k$ represents the total
distance, in reciprocal space and in units of $2\pi /a_{0}$ along the path $%
\Gamma -J-X-\Gamma $ from the origin $\Gamma $. The legend indicates in what
response function $\chi _{\rho _{n},\rho _{n}},\chi _{s_{x},s_{x}},\chi
_{s_{y},s_{y}},$ or $\chi _{s_{z},s_{z}}$ the collective mode has the
biggest weight. This gives an indication of the nature of the mode. In Fig. %
\ref{fig4}, the phonon mode has its biggest weight in $\chi _{\rho _{n},\rho
_{n}}$ and $\chi _{S_{x},S_{x}}$ while the pseudospin wave mode has its
weight in $\chi _{S_{y},S_{y}}$ and $\chi _{S_{z},S_{z}}$. That, is, since
we forced the pseudospin to be polarized along the $x$ direction, the
pseudospin wave mode corresponds to a precession of the pseudospin about the 
$x$ axis. The phonon dispersion is typical of what is found for a Wigner
crystal\cite{cotemethode}. It is gapless, with $\omega \sim q^{3/2}$
behavior at small wave vector. For the pseudospin wave, the dispersion is $%
\omega \sim q^{2}$ at small wave vector confirming the SU(2) symmetry. At $%
\nu _{0}=0.2,$ the bandwidth of the pseudospin mode is two orders of
magnitude smaller than that of the phonon mode. While the bandwidth of the
phonon mode does not change much as $\nu _{0}$ increases to $\nu _{0}=0.5$,
that of the pseudospin mode changes dramatically, becoming of the same order
as that of the phonon mode at $\nu _{0}=0.5.$ The pseudospin stiffness thus
increases rapidly with $\nu _{0}.$ The hBC1 dispersion has the same features
as the eBC1 as can be seen in Fig. \ref{fig5}.

\begin{figure}[tbph]
\includegraphics[scale=1]{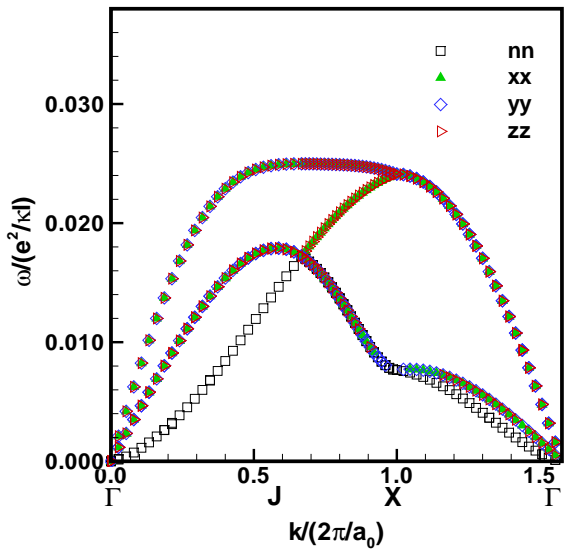}
\caption{Dispersion relation of the Goldstone modes of the MPC state at $%
\protect\nu _{0}=0.8$ in Landau level $n=0.$ The dispersion is plotted along
the irreducible Brillouin zone of the triangular lattice. The legend
indicates in what response function each collective mode has its biggest
weight.}
\label{fig6}
\end{figure}

\begin{figure}[tbph]
\includegraphics[scale=1]{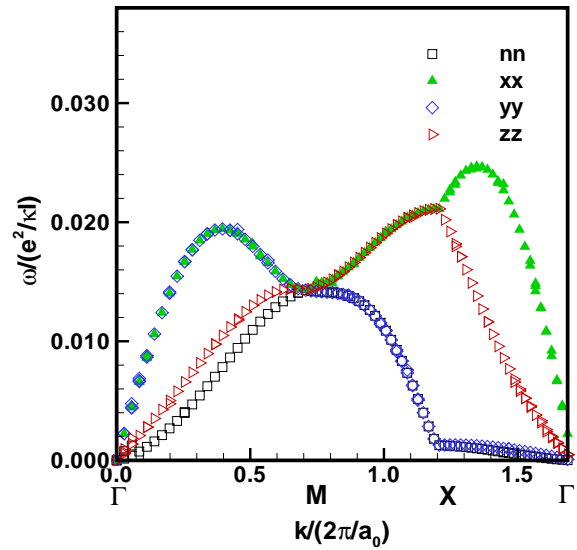}
\caption{Dispersion relation of the Goldstone modes of the MC state at $%
\protect\nu _{0}=0.625$ in Landau level $n=0.$ The dispersion is plotted
along the irreducible Brillouin zone of the square lattice. The legend
indicates in what response function each collective mode has its biggest
weight.}
\label{fig7}
\end{figure}

For $\nu >0.55$ in $n=0$, we find that, within our numerical accuracy, the
MC and MPC\ have the same energy and are lower in energy than the other
phases considered. The dispersion relations, however, indicate that the MC\
is stable in the range $\nu \in \left[ 0.55,0.65\right] $ while the MPC\ is
unstable in that range and vice versa for $\nu \in \left[ 0.65,0.90\right] $
so that there is a phase transition between these two states. We show in
Figs. \ref{fig6} and \ref{fig7} the dispersion relations for these two
states. For the MC phase, the dispersion is plotted along the path $\Gamma
-M-X-\Gamma ,$corresponding to the wave vectors $\left( k_{x},k_{y}\right)
=\left( 0,0\right) ,\left( 2\pi _{0}/a\right) \left( 1/2,1/2\right) ,\left(
2\pi /a_{0}\right) \left( 1/2,0\right) ,\left( 0,0\right) $ since the unit
cell is that of a square lattice. The dispersion in both cases show the
usual gapless phonon mode with $\omega \sim q^{3/2}$ behavior at small wave
vector which appears as a pole of $\chi _{\rho _{n},\rho _{n}}$ and $3$
other linearly dispersing Goldstone modes. Some of the pseudospin modes are
degenerate along sections of the contour of the irreducible Brillouin zone.
The degeneracy of some of these modes is only lifted along $\Gamma -X$ in
the MC\ phase in \ref{fig7}. For wave vector $\mathbf{k}$ in an arbitrary
direction, however, the $3$ pseudospin modes are non degenerate.

The energy of the two meron lattices are invariant under a rotation of the
pseudospin texture around the $x,y$ or $z$-axis since there is no equivalent
of the Zeeman coupling in the graphene 2DEG. This implies there are three
independent ways to rotate the spins of the state without any cost in
energy, leading to the three Goldstone modes found in the GRPA. Animations
of these three modes support this interpretation\cite{cotecp3}.

\begin{figure}[tbph]
\includegraphics[scale=1]{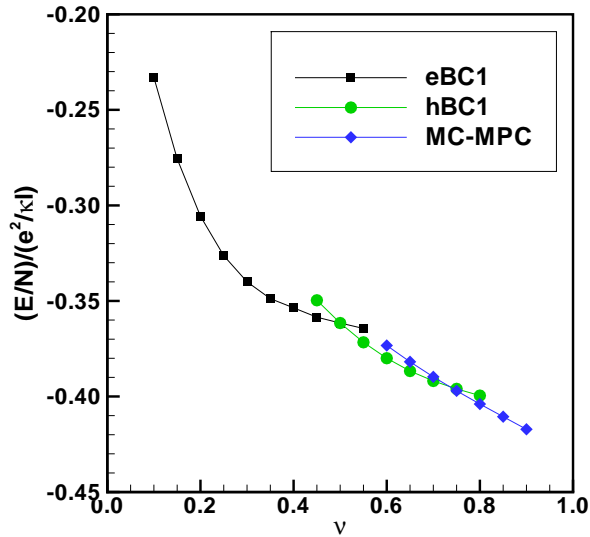}
\caption{Hartree-Fock energy per electron as a function of filling factor
for various crystal phases in Landau level $n=1.$}
\label{fig8}
\end{figure}

We show in Fig. \ref{fig8} the energies of different phases of the 2DEG in
graphene for Landau level $n=1.$ We find the following sequence: eBC1 for $%
\nu _{1}\in \left[ 0.1,0.5\right] ,$ hBC1 for $\nu _{1}\in \left[ 0.5,0.75%
\right] ,$ MC for $\nu _{1}\in \left[ 0.75,0.80\right] $ and MPC for $\nu
_{1}\in \left[ 0.75,0.90\right] $. The MC\ and MPC phases have almost the
same energy within our numerical accuracy so that these two phases are
represented by the line MC-MPC\ in \ref{fig8}. In comparison with the case $%
n=0$, we see that the filling factor range for which a pseudospin texture
exists for $n=1$ has decreased relative to $n=0$. In contrast with what
happens in a conventional 2DEG, however, there is a possibility for such
textures in Landau level $n=1.$ This is due to the fact that the effective
interactions in the two systems are now different in view of Eq. (\ref{2_9}).

The dispersion of the gapless modes in Landau levels $n=1$ and $n=2$ are
similar to what is seen in $n=0$ as can be seen from Fig. \ref{fig5}. Bubble
crystals with more than one electron per site have additional gapped modes
related to internal excitations of the bubbles; we do not focus on these
modes in this work\cite{cotebubble}.

\begin{figure}[tbph]
\includegraphics[scale=1]{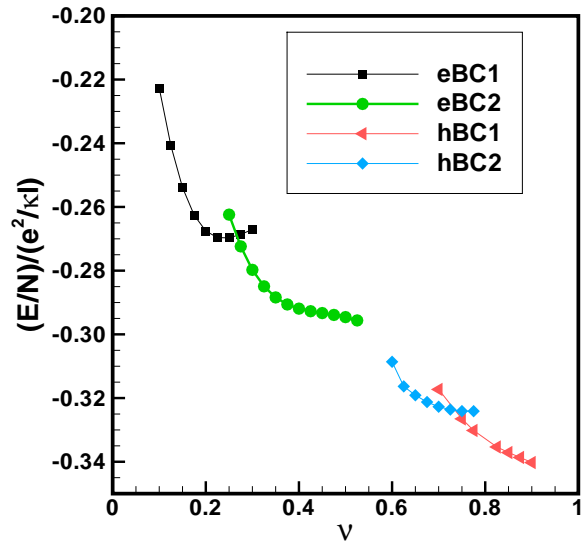}
\caption{Hartree-Fock energy per electron as a function of filling factor
for various crystal phases in Landau level $n=2.$}
\label{fig9}
\end{figure}

Fig. \ref{fig9} shows the phase diagram of the 2DEG in graphene for Landau
level $n=2.$ We find here a sequence of transitions involving electron and
hole bubble crystals with one or two electrons per bubble. The two meron
phases MC\ and MPC\ have higher energy than the other phases considered so
that there are no meron crystals of these types in $n=2$ and most probably
in higher Landau levels as well. We emphasized however that our calculation
is restricted to the range of partial filling factor $\nu -1>0.1$. Because
the pseudospin textured states are pushed closer to $\nu =1$ as $n$
increases, we cannot rule the existence of meron crystals for $n\geq 2.$
Indeed, Yang, Das Sarma and MacDonald have shown that skyrmions are the
lowest-energy charged excitations in graphene for Landau levels up to $n=3$%
\cite{macdoyan}. Meron crystals could thus also be present in Landau levels $%
n=2,3$.

For $n=2$ and near filling factor $\nu _{2}=0.5$, none of the phases that we
considered in our analysis are stable so that the ground state must be of
another crystal type, most probably the stripe phase if we compare with the
situation in semiconductor 2DEG. This is also the conclusion of Hartree-Fock
calculations in Refs. \onlinecite{joglekar} and \onlinecite{jianhui}. A
stripe state probably occurs near $\nu _{n}=0.5$ for Landau level $n\geq 2$
in the HFA.

\section{Signatures of Wigner and meron crystals in pinning behavior}

Generally, collective modes of 2DEG's are detected by inelastic light
scattering or via microwave absorption. The latter experiments are most
sensitive to the long wavelength, low frequency behavior of the collective
modes. A disorder potential pins the Wigner crystal, in the sense that the
phonon mode becomes gapped at a pinning frequency $\omega _{p}$ that is
dependent on the strength of the potential and on that of the magnetic
field. The behavior of the pinning frequency with magnetic field depends
critically on the interplay between different length scales: the size of the
electron wavefunction on each lattice site, the lattice periodicity, and the
magnetic length\cite{chitra,fertigpin}. The pinning frequency in the
longitudinal conductivity may be inferred from results such as those found
here using the replica trick\cite{pinnedbubble}.

In the graphene case, both the pseudospin and phonon mode involve charge
fluctuations so we may speculate that the pseudospin mode will also be
pinned in the presence of disorder, in the sense of opening a gap in their
spectrum, because the latter generically breaks pseudospin symmetry.
Disorder should also pin the four Goldstone modes of the meron crystals: the
phonon mode and the three pseudospin wave modes. If such pinned modes are
separately observable, they could provide a unique signature of the
formation of a Wigner or meron crystals in graphene. This conclusion should
be contrasted with the case of a semiconductor 2DEG. There, a Wigner crystal
has only one gapless (phonon) mode at finite Zeeman coupling. A skyrmion
(finite Zeeman coupling) or meron (zero Zeeman coupling)\ crystal has one
phonon mode which is gapped and one (Wigner)\ or three (meron) gapless spin
wave modes that remain gapless in the presence of disorder.

For the pinning modes to be visible in microwave absorption experiments\cite%
{microwave}, they must show up in $\Re\left[ \sigma _{xx}\left( \mathbf{k}%
\rightarrow 0,\omega \right) \right] $ where $\sigma $ is the conductivity
tensor. Equivalently, they must appear as poles of the current-current
response functions $\Im\left[ \chi _{i,i}^{J,J}\left( \mathbf{k}\rightarrow
0,\omega \right) \right] $ with $i=x,y.$ A calculation of the conductivity
tensor in the presence of disorder is difficult for the crystal states. For
a simple Wigner crystal (one phonon mode only), it can be done by mapping
the system to an effective harmonic model\cite{chitra,pinnedbubble} and
using the replica trick. So far there has been no generalization of this
method to crystals with an additional layer or valley degree of freedom. To
test our speculation that the phonon and valley pseudospin modes are visible
in microwave absorption, we \textquotedblleft simulate\textquotedblright\ a
disorder potential by adding a periodic external potential. Note that our
HFA method forces us to choose this periodicity to be the same as that of
the crystal considered.

In formulating the relevant response function for conductivity, the current
operator is built up from operators that excite electrons between Landau
levels. This presents problems when the Hilbert space is restricted to one
Landau level as in our calculation. In principle one needs to retain other
Landau levels in order to obtain a non-vanishing result, significantly
complicating the calculations. To circumvent this, we wish to find an
appropriate projection of the current operator into a single Landau level.
To do this, we generalize a method first introduced by Girvin, MacDonald and
Platzman in Ref. \onlinecite{currentgirvin}. This procedures captures the
drift current $\mathbf{j}\left( \mathbf{r}\right) =\frac{ce}{B}\rho \left( 
\mathbf{r}\right) \nabla U\left( \mathbf{r}\right) \times \widehat{\mathbf{z}%
}$ of the electrons in the potential $U\left( \mathbf{r}\right) $ and gives
a current that satisfies the continuity equation. Improvements upon this
procedure are possible but lead to very complicated expressions\cite{current}%
. We first write a second quantized hamiltonian including the (total)
density $\rho \left( \mathbf{k}\right) $ and valley-pseudospin operator $%
\mathbf{P}\left( \mathbf{k}\right) $

\begin{eqnarray}
H &=&N_{\varphi }\frac{1}{4}\left( \frac{e^{2}}{\kappa \ell }\right) \sum_{%
\mathbf{q}}\Upsilon \left( \mathbf{q}\right) \rho \left( -\mathbf{q}\right)
\rho \left( \mathbf{q}\right)  \label{hamil} \\
&&-N_{\varphi }\left( \frac{e^{2}}{\kappa \ell }\right) \sum_{\mathbf{q}%
}X\left( \mathbf{q}\right) \mathbf{P}\left( -\mathbf{q}\right) \cdot \mathbf{%
P}\left( \mathbf{q}\right) ,  \notag
\end{eqnarray}%
where 
\begin{equation}
\Upsilon \left( \mathbf{q}\right) =2H\left( \mathbf{q}\right) -X\left( 
\mathbf{q}\right) .
\end{equation}%
With the hamiltonian of Eq. (\ref{hamil}), we obtain the equation of motion
of the density operator%
\begin{equation}
i\hslash \frac{d\rho \left( \mathbf{k}\right) }{dt}=\left[ \rho \left( 
\mathbf{k}\right) ,H\right] ,
\end{equation}%
which we linearize by writing $\rho \left( \mathbf{k}\right) \rightarrow
\left\langle \rho \left( \mathbf{k}\right) \right\rangle +\delta \rho \left( 
\mathbf{k}\right) $ where the average is evaluated in the HFA. Keeping terms
up to linear order in $\delta \rho \left( \mathbf{k}\right) ,$we find 
\begin{eqnarray}
\hslash \frac{d\delta \rho \left( \mathbf{k}\right) }{dt} &=&\left( \frac{%
e^{2}}{\kappa \ell }\right) \sum_{\mathbf{G}}\left[ \Upsilon \left( \mathbf{G%
}\right) -\Upsilon \left( \mathbf{k-G}\right) \right] \\
&&\times \sin \left( \mathbf{k}\times \mathbf{G}\ell ^{2}/2\right)
\left\langle \rho \left( \mathbf{G}\right) \right\rangle \delta \rho \left( 
\mathbf{k-G}\right)  \notag \\
&&-4\left( \frac{e^{2}}{\kappa \ell }\right) \sum_{\mathbf{G}}\left[ X\left( 
\mathbf{G}\right) -X\left( \mathbf{k-G}\right) \right]  \notag \\
&&\times \sin \left( \mathbf{k}\times \mathbf{G}\ell ^{2}/2\right)
\left\langle \mathbf{P}\left( \mathbf{G}\right) \right\rangle \cdot \delta 
\mathbf{P}\left( \mathbf{k-G}\right) .  \notag
\end{eqnarray}%
This is the equation of motion of the density in the GRPA.

To find an expression for the current operator valid at small $\mathbf{k}$,
we make the approximation 
\begin{equation}
\sin \left( \mathbf{k}\times \mathbf{G}\ell ^{2}/2\right) \approx \mathbf{%
k\cdot }\left( \mathbf{G}\times \widehat{\mathbf{z}}\ell ^{2}/2\right) ,
\end{equation}%
and use the continuity equation 
\begin{equation}
\frac{d\delta \rho \left( \mathbf{k}\right) }{dt}=\frac{i}{e}\mathbf{k}\cdot 
\mathbf{j}\left( \mathbf{k}\right) .
\end{equation}%
We get in this way 
\begin{eqnarray}
\mathbf{j}\left( \mathbf{k}\right) &=&-\frac{i}{2}\left( \frac{e^{3}}{%
\hslash \kappa }\right) \sum_{\mathbf{G}}\left( \mathbf{G}\times \widehat{%
\mathbf{z}}\ell \right) \left[ \Upsilon \left( \mathbf{k+G}\right) -\Upsilon
\left( \mathbf{G}\right) \right]  \notag \\
&&\times \left\langle \rho \left( -\mathbf{G}\right) \right\rangle \delta
\rho \left( \mathbf{k+G}\right)  \label{currentpp} \\
&&+2i\left( \frac{e^{3}}{\hslash \kappa }\right) \sum_{\mathbf{G}}\left( 
\mathbf{G}\times \widehat{\mathbf{z}}\ell \right) \left[ X\left( \mathbf{k+G}%
\right) -X\left( \mathbf{G}\right) \right]  \notag \\
&&\times \left\langle \mathbf{P}\left( -\mathbf{G}\right) \right\rangle
\cdot \delta \mathbf{P}\left( \mathbf{k+G}\right) .  \notag
\end{eqnarray}%
This expression shows that the current can have contributions from both
density and pseudospin fluctuations. In terms of the original $\rho
_{i,j}\left( \mathbf{q}\right) $ operators, we can write the current
expression as%
\begin{equation}
\mathbf{j}\left( \mathbf{k}\right) =-i\frac{e^{3}}{\hslash \kappa }\sum_{%
\mathbf{G}}\sum_{a,b}F_{a,b}\left( \mathbf{k},\mathbf{G}\right) \delta \rho
_{a,b}\left( \mathbf{k}+\mathbf{G}\right) ,  \label{current}
\end{equation}%
where 
\begin{eqnarray}
&&\mathbf{F}_{a,b}\left( \mathbf{k},\mathbf{G}\right) \\
&=&\left( \mathbf{G}\ell \times \widehat{\mathbf{z}}\right) \left[ H\left( 
\mathbf{k}+\mathbf{G}\right) -H\left( \mathbf{G}\right) \right] \left\langle
\rho \left( -\mathbf{G}\right) \right\rangle \delta _{a,b}  \notag \\
&&-\left( \mathbf{G}\ell \times \widehat{\mathbf{z}}\right) \left[ X\left( 
\mathbf{k}+\mathbf{G}\right) -X\left( \mathbf{G}\right) \right] \left\langle
\rho _{b,a}\left( -\mathbf{G}\right) \right\rangle .  \notag
\end{eqnarray}

With Eq. (\ref{current}) for the current, we can easily write the
current-current Matsubara Green's function tensor 
\begin{equation}
\chi ^{J,J}\left( \mathbf{k},\tau \right) =-N_{\varphi }\left\langle T%
\mathbf{j}\left( \mathbf{k},\tau \right) \mathbf{j}\left( -\mathbf{k}%
,0\right) \right\rangle ,
\end{equation}%
so that the retarded current-current response function is finally given by 
\begin{eqnarray}
\chi ^{J,J}\left( \mathbf{k},\omega \right) &=&-\left( \frac{e^{3}}{\hslash
\kappa }\right) ^{2}\sum_{a,b,c,d}\sum_{\mathbf{G},\mathbf{G}^{\prime }}%
\mathbf{F}_{a,b}\left( \mathbf{k},\mathbf{G}\right)  \label{current1} \\
&&\times \chi _{n}^{a,b,c,d}\left( \mathbf{k}+\mathbf{G},\mathbf{k}+\mathbf{G%
}^{\prime },\omega \right) \mathbf{F}_{c,d}\left( -\mathbf{k},-\mathbf{G}%
^{\prime }\right) .  \notag
\end{eqnarray}

If we apply an external potential, the hamiltonian $H\rightarrow H+H_{ext}$
with 
\begin{equation}
H_{ext}=N_{\phi }\left( \frac{e^{2}}{\kappa \ell }\right) \sum_{a}\sum_{%
\mathbf{q}}W_{a}\left( -\mathbf{q}\right) \rho _{a,a}\left( \mathbf{q}%
\right) .  \label{coupling}
\end{equation}
We allow the potential $W_{a}\left( \mathbf{q}\right) $ to be different for
the $K$ and $K^{\prime }$ valleys. With this potential, the function $%
\mathbf{F}_{a,b}\left( \mathbf{k},\mathbf{G}\right) $ in Eq. (\ref{current1}%
) must be replaced by

\begin{eqnarray}
&&\mathbf{F}_{i,j}\left( \mathbf{k},\mathbf{G}\right)  \label{ffunction} \\
&=&-\left( \mathbf{G}\times \widehat{\mathbf{z}}\ell \right) W_{K}\left( -%
\mathbf{G}\right) \delta _{i,j}\delta _{i,K}  \notag \\
&&-\left( \mathbf{G}\times \widehat{\mathbf{z}}\ell \right) W_{K^{\prime
}}\left( -\mathbf{G}\right) \delta _{i,j}\delta _{i,K^{\prime }}  \notag \\
&&+\left( \mathbf{G}\ell \times \widehat{\mathbf{z}}\right) \left[ H\left( 
\mathbf{k}+\mathbf{G}\right) -H\left( \mathbf{G}\right) \right] \left\langle
\rho \left( -\mathbf{G}\right) \right\rangle \delta _{i,j}  \notag \\
&&-\left( \mathbf{G}\ell \times \widehat{\mathbf{z}}\right) \left[ X\left( 
\mathbf{k}+\mathbf{G}\right) -X\left( \mathbf{G}\right) \right] \left\langle
\rho _{j,i}\left( -\mathbf{G}\right) \right\rangle .  \notag
\end{eqnarray}%
In pseudospin language, this means that the current in Eq. (\ref{currentpp})
becomes $\mathbf{j}\left( \mathbf{k}\right) \rightarrow \mathbf{j}\left( 
\mathbf{k}\right) +\mathbf{j}_{W}\left( \mathbf{k}\right) ,$ where%
\begin{eqnarray}
\mathbf{j}_{W}\left( \mathbf{k}\right) &=&-\frac{i}{4}\left( \frac{e^{3}}{%
\hslash \kappa }\right) \sum_{\mathbf{G}}\left( \mathbf{G}\times \widehat{%
\mathbf{z}}\ell \right) W_{+}\left( -\mathbf{G}\right) \delta \rho \left( 
\mathbf{k+G}\right) \\
&&-\frac{i}{2}\left( \frac{e^{3}}{\hslash \kappa }\right) \sum_{\mathbf{G}%
}\left( \mathbf{G}\times \widehat{\mathbf{z}}\ell \right) W_{-}\left( -%
\mathbf{G}\right) \delta P_{z}\left( \mathbf{k+G}\right) ,  \notag
\end{eqnarray}%
where $W_{\pm }\left( -\mathbf{G}\right) =W_{K}\left( -\mathbf{G}\right) \pm
W_{K^{\prime }}\left( -\mathbf{G}\right) .$ Note that for consistency, we
also include the external potential in the calculation of the $\left\langle
\rho _{i,j}\left( \mathbf{G}\right) \right\rangle ^{\prime }s$ as well as in
that of $\chi _{n}^{a,b,c,d}\left( \mathbf{k}+\mathbf{G},\mathbf{k}+\mathbf{G%
}^{\prime },\omega \right) .$

In the absence of any external potential, we find that, for the Wigner or
meron crystal, only the phonon mode (and some higher energy modes)\ appears
as a pole of $\Im\left[ \chi _{x,x}^{J,J}\left( \mathbf{k},\omega \right) %
\right] $ for any wave vector $\mathbf{k}$. The phase modes are
conspicuously absent of the current response. We explain this by the fact
that the phase modes are transverse modes: the motion of a pseudospin is
perpendicular to the local value of that pseudospin so that the term $%
\left\langle \mathbf{P}\left( -\mathbf{G}^{\prime }\right) \right\rangle
\cdot \delta \mathbf{P}\left( \mathbf{k+G}^{\prime }\right) =0$ in Eq. (\ref%
{currentpp}). Also, these modes have no weight in the density-density
response function. They cannot contribute to the local current.

\begin{figure}[tbph]
\includegraphics[scale=1]{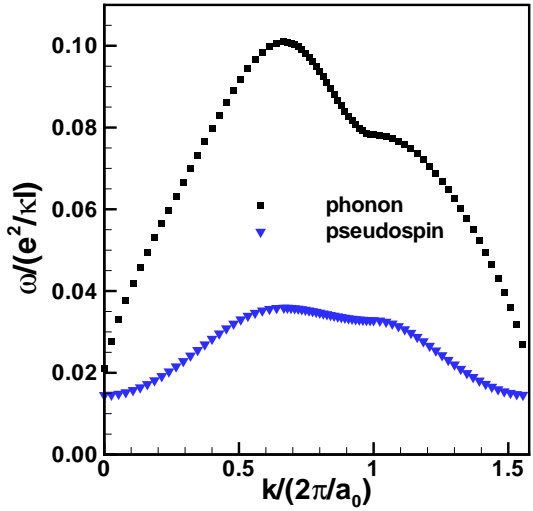}
\caption{Dispersion relation of the phonon and pseudospin modes of the
Wigner crystal at $\protect\nu _{0}=0.4$ in Landau level $n=0$ in an
external potential $W_{K}=-0.005$ and $W_{K^{\prime }}=0.$ The dispersion is
plotted along the irreducible Brillouin zone of the triangular lattice (see
Fig. 5). Both the phonon and pseudospin wave mode are gapped by the external
field.}
\label{fig10}
\end{figure}

This conclusion is unchanged, \textit{for the Wigner crystal state}, if we
apply an external potential with $W_{K}\left( \mathbf{q}\right)
=W_{K^{\prime }}\left( \mathbf{q}\right) .$ The phonon mode is gapped by
that potential but the pseudospin mode dispersion remains gapless because
the pseudospin symmetry has not been broken. To induce a gap in the
pseudospin mode, we must allow $W_{K}\left( \mathbf{q}\right) \neq
W_{K^{\prime }}\left( \mathbf{q}\right) $ so that the external potential can
couple to the $z$ component of the pseudospin: $P_{n,z}\left( \mathbf{q}%
\right) =(\rho _{n}^{+,+}\left( \mathbf{q}\right) -\rho _{n}^{-,-}\left( 
\mathbf{q}\right) )/2.$ We show in Fig. 10 the dispersion relation of the
collective modes of the Wigner crystal for $\nu =0.4$ and with an external
potential that is different in the two valleys i.e. $W_{K}=-0.005$ and $%
W_{K^{\prime }}=0$. Both the phase and phonon modes are now gapped as
expected but Fig. 11 shows that, once again, only the phonon appears as pole
of the current. This fact can be readily understood: the external potential
is much weaker than the exchange energy that forces the parallel alignment
of the pseudospins. It follows that when the external potential is applied,
the pseudospins \textit{all} align along the $z$ axis even though the
external potential is modulated in space. The pseudospin mode is a
transverse mode and so both terms $\left\langle \mathbf{P}\left( -\mathbf{G}%
^{\prime }\right) \right\rangle \cdot \delta \mathbf{P}\left( \mathbf{k+G}%
^{\prime }\right) $ and $\delta P_{z}\left( \mathbf{k+G}\right) $ in the
definition of the current in Eq. (\ref{currentpp}) are zero. For the phonon
mode in the Wigner crystal state, the peak in the current response stays
finite with decreasing wave vector so that the phonon mode is visible at $%
\mathbf{k}=0$ and can contribute to the absorption.

\begin{figure}[tbph]
\includegraphics[scale=1]{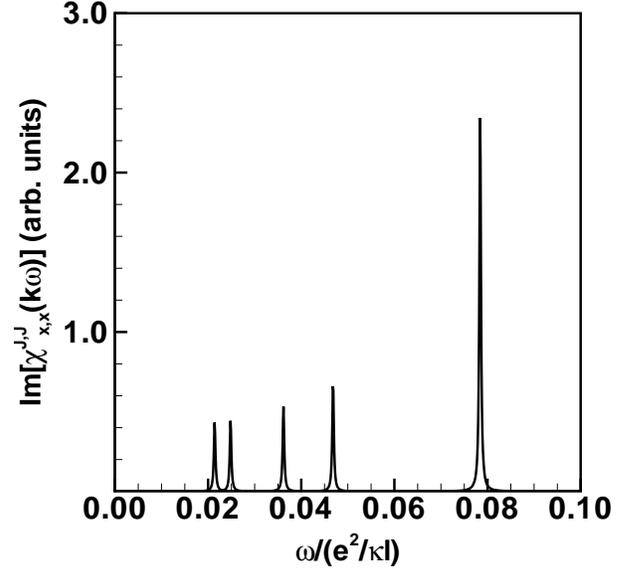}
\caption{Imaginary part of the current response function. From left to
right: $k_{x}=k_{y}=0.001,0.01,0.05,0.1,0.3$ in units of $2\protect\pi %
/a_{0}.$ These peaks come from the phonon mode. There is no peak coming from
the pseudospin mode.}
\label{fig11}
\end{figure}

For the meron crystal, the situation is more complicated. An external
potential $W_{-}\left( -\mathbf{G}\right) $ acts as a pseudomagnetic field.
If it is uniform in space, it acts effectively as a pseudospin Zeeman
coupling and produces a transition from a meron to a bimeron crystal with
two Goldstone modes (the phonon mode and a gapless pseudospin mode related
to the $U(1)$ symmetry of the hamiltonian) and two gapped pseudospin modes%
\cite{cotecp3}. The pseudospin modes being transverse and the second term in 
$\mathbf{j}_{W}\left( \mathbf{k}\right) $ being zero because of the
uniformity of the potential, there should again be no contribution of these
modes to the conductivity. This is the case for two of the pseudospin modes,
but the gapless pseudospin mode has a weight in the density-density response
function at finite wave vector $\mathbf{k}$ and does appear in the current
response along with the phonon peak. Both peaks go to zero with decreasing
wave vector however.

For the meron crystal, the second term in $\mathbf{j}_{W}\left( \mathbf{k}%
\right) $ should be finite if $W_{-}\left( -\mathbf{G}\right) $ is
non-uniform, in contrast with the Wigner crystal case, because some of the
pseudospin modes involve a fluctuation in the $\hat{z}$ component of the
pseudospin. Unfortunately, we find the meron crystal to be very sensitive to
an inhomogeneous external potential, as indicated by instabilities in the
collective mode spectrum at small values of $k$. This is likely related to
the extreme closeness in energy of the meron crystal and meron pair crystal
states, so that the external potential may lead to a different and possibly
more complicated textured state. Thus it is not possible to compute the
response functions at small wave vector $k$ for these textured states
without better knowledge of their groundstate structure in the presence of a
pinning potential. We note that at finite wavevector, where the dispersion
is well behaved, the pseudospin modes do appear in the current response as
expected.

In spite of the difficulty demonstrating the presence of a signature of the
pinned phase modes in the dynamical conductivity at small wavevector, we
believe at least a small response will in fact generically always be
present. Beyond the density fluctuations in the pseudospin modes due to
spin-charge coupling, there is a further density response due to the fact
that the $A$ and $B$ sites of the lattice are at difference positions in
real space in a unit cell. In our calculations, $A$ and $B$ were treated as
two orthogonal \textquotedblleft spin\textquotedblright\ states of an
electron, but their slightly different locations in real space were not
included in the model. If this were included, we expect that an oscillation
of the pseudospin that changes the relative weight of an electron on the $A$
or $B$ sublattice (or, equivalently, on the $K$ and $K^{\prime }$ valleys)
will translate into a change in the position of that electron or into a
dipole fluctuation. Thus, if this distinction were properly included in our
model, we would expect that the phase mode would appear as a pole of the
current response just as the phonon mode does, albeit weakly, since the
symmetry breaking is small.

In closing this section, we remark that, in Fig. \ref{fig4}, the effective
stiffness for the pseudospin mode is two orders of magnitude smaller than
that of the Wigner crystal for filling factor $\nu \lesssim 0.2$. In the
presence of disorder, we might then expect two pinning modes of very
different frequencies for a Wigner crystal and it may be impossible to
detect the two modes simultaneously in an actual experiment. The meron
crystal dispersion does not suffer from this problem since all four modes
appear to have similar bandwidths.

\section{Conclusion}

We have shown in this work that the 2DEG in graphene can support Wigner
crystals and meron crystals with valley-pseudospin textures. Our numerical
analysis was restricted to filling factor $\nu _{n}\in \left[ 0.1,0.9\right] 
$ in each Landau level $n$ and we concluded that, in this range, meron
crystals are present in Landau levels $n=0$ and $n=1$ only. We have computed
the dispersion relation of the collective excitations of these two crystal
states and showed that the Wigner crystal has one extra Goldstone mode with
a quadratic dispersion at small wave vector in addition to the phonon mode.
Meron crystals have 3 extra Goldstone modes in addition to the phonon mode.
These extra Goldstone modes are valley-pseudospin fluctuations. In graphene,
all these modes involve density fluctuations, and we speculated that these
last modes could be visible as pinning modes in microwave absorption
spectrum in a real disordered system.

\begin{acknowledgments}
This work was supported by a research grant from the Natural Sciences and
Engineering Research Council of Canada (NSERC) for R. C\^{o}t\'{e} and an
NSF Grant No. DMR-0704033 for H. A. Fertig. Computer time was provided by
the R\'{e}seau Qu\'{e}b\'{e}cois de Calcul Haute Performance (RQCHP).
\end{acknowledgments}

\end{document}